\documentstyle[aps,eqsecnum,preprint]{revtex}
\begin{document}
\draft
\title{Molecular theory of elastic constants of liquid crystals. III.
Application to smectic phases with tilted orientational order}
\author{ Yashwant Singh and Jokhan Ram}
\address{ Department of Physics, Banaras Hindu University, \\
     Varanasi 221 005, India }
\date{\today}
\maketitle
\begin{abstract}

Using the density functional formalism we derive expression for the 
distortion free energy for systems with continuous broken symmetry 
and use it to derive expression for the elastic constants of smectic 
phases in which director is tilted with respect to the smectic layer 
normal. As in the previous papers of the series ( Phys. Rev. A {\bf 
45}, 974 (1992), E {\bf 49}, 501, (1994) ) the expressions for the 
elastic constants are written in terms of order and structural 
parameters. The structural parameters involve  the generalised spherical 
harmonic coefficients of the direct pair correlation function of an 
effective isotropic liquid. The density of this effective isotropic 
liquid depends on the nature and amount of ordering present in the 
system and is evaluated self- consistently.  We estimate the value of 
elastic constants using reasonable guess for the order and structural-
parameters.
\end{abstract}                                                               
\narrowtext 

\section{Introduction}

In a previous paper \cite{1} of this series ( hereafter referred 
to as I ) we developed a theory based on the density functional 
formalism \cite{2} for the deformation free energy of any system 
with continuous broken symmetry and applied the theory to derive
expressions for the elastic constants of uniaxial nematic and 
smectic A (Sm A) phases. In the second paper \cite{3} (hereafter is 
referred to as II) the theory was applied to derive expression for 
all the twelve bulk elastic constants of biaxial nematic phase of 
orthorhombic symmetry.  These expressions of  elastic 
constants are written in terms of order parameters which characterize
the nature and amount of ordering in the phase and the structural 
parameters  which involve the generalized spherical harmonic 
coefficients of the direct pair correlation function of an effective 
isotropic liquid the density of which is determined using a criterion 
of the weighted density-functional formalism. The purpose of this paper 
is to expand upon the theory given in I and apply it to those smectic phases 
in which the director is tilted with respect to the smectic layer normal.
This class of smectics include Sm C, Sm I, Sm F and Sm K phase and their 
chiral versions. 

In general, smectic liquid crystals have a stratified structure with the
long axes of the rod-like or lath-like molecules parallel to each other
in the layers. This situation corresponds to partial breakdown of 
translational invariance. Since a variety of molecular arrangements
are possible within each layer a number of smectic phases are possible.
Thus a smectic is defined by its periodicity in one direction of space
and by its point group symmetry which is continuous or discrete 
subgroup of $D_{\infty h}$. In this article we are confined to those
smectic phases which point group symmetries are $C_{2h}$ and $C_2$.
Most important phases of this class of smectics as far as elasticity
is concerned, are Sm C and Sm${\rm C}^*$ phases. 

In Sm C the molecules are arranged as shown in Fig.1. Each smectic layer
is a two dimensional liquid. The director $\hat{\bf n}$ makes a finite
angle with the normal to the layer. The tilted orientational order is
described by a two dimensional unit vector which is along the 
projection of $\hat{\bf n}$
on the layer and is denoted as $\hat{\bf c}$ ( see Fig.1). 
The direction of ${\bf \hat c}$ is defined
relative to a chosen axis in the plane and simultaneously
with the sign of projection of $\hat{\bf n}$ on the normal
to the layer. The correlation
of the tilt direction of different layers implies that $\hat{\bf c}$ must 
be uniform over macroscopic distance. Though Sm C is $\hat{\bf n} 
\leftrightarrow -\hat{\bf n}$ symmetric, such symmetry operation does not 
hold for the  $\hat{\bf c}$ -director. The states described by  
$\hat{\bf c}$ and  $-\hat{\bf c}$ are not equivalent and hence 
$\hat{\bf c}$ is a two dimensional polar vector. The tilted structure 
introduces biaxiallity within the layers and Sm C is optically biaxial 
with point symmetry $C_{2h}$.

The other non-chiral smectic phases of this class namely Sm I, Sm F 
and Sm K all possess more order than the Sm C phase. In these each smectic
layer is an independent two dimensional bond orientational ordered system.
The coupling between neighbouring hexatic layers drives the quasi long 
range hexatic bond-orientational ordering of a layer into a truly long range
ordered hexatic state. Therefore these phases have long ranged 
bond-orientational order but short ranged positional order in the layer
in a three dimensional stacked hexatic systems. 
The director $\hat{\bf n}$, as in Sm C, makes a finite angle with 
the layer normal. These phases, in spite
of having quite different local order, have the same symmetry as a Sm C
phase. The phase transition between Sm C and one of the tilted hexatic
phases is found to be first order. This transition is analogous to the 
3-dimensional liquid to vapor transition  which is first order upto a 
critical point, and beyond the critical point there is no real phase 
transition. One can go continuously from one to the other phase 
without phase transition at all. The expression for the elastic free 
energy density for these phases are, therefore, similar to that of Sm C;
the difference can be only quantitative. In view of this we discuss the
elasticity of Sm C phase only.

The chiral smectic C (Sm${\rm C}^*$) has modulated structure at a scale
dimension of the order of 1 $\mu$m and larger. The modulated
(helical) structure occurs as a result of a precession of the molecular
tilt about an axis perpendicular to the layer planes. The tilt direction
is rotated through an azimuthal angle $\phi$ on moving from one layer
to the next. As the rotation is in a constant direction, a helix
is formed which is either left or right handed. In Sm${\rm C}^*$ the mirror
plane is lost but $C_2$ axis is present locally in each smectic layer
(it is perpendicular to the plane containing the molecular axes).
The $C_2$ axis in the case of  Sm${\rm C}^*$ phase is a polar axis which
admits the existence of a spontaneous polarization along it. The 
Sm${\rm C}^*$  phase  therefore exhibits ferroelectricity.

Many compounds exhibit phases such as antiferroelectric (Sm${\rm C_A^*}$), 
ferroelectric (Sm ${\rm C_r^*}$) etc in addition to  Sm${\rm C}^*$, 
as the temperature is varied. 
The director $\hat{\bf n}$ in the Sm${\rm C_A^*}$ 
phase, like in Sm${\rm C}^*$ phase is uniformly tilted with respect to 
the layer normal. However, unlike in Sm${\rm C}^*$ phase, the difference 
in azimuthal angle of {\bf n} in successive layer is $\pi + \alpha$. 
The small angle $\alpha$ gives rise to helical structure of {\bf n} along
the layer normal and arises from the chirality of the constituent 
molecules. In Sm${\rm C_r^*}$ the layers are stacked in such a way that there
is a net overall (in contrast to Sm${\rm C_A^*}$ phase) spontaneous
polarization. This means the number of layers of opposite polarization
is not equal. The elasticity associated with the structure of all 
chiral smectic phases is same as that of Sm${\rm C}^*$  phase. 

The paper is organised as follows : 
In Sec. II we describe the elastic continuum theory 
for Sm C and Sm${\rm C^*}$ phases and derive expression for the elastic
free energy density. This expression is then written in terms of 
deformation variables so that it becomes appropriate for comparison
with an expression  found from microscopic theory. The
density functional approach has been used in sec. III to describe the 
deformed state of smectics with tilted orientational order. The expression
for the deformation free energy density derived here is more general
than that given in I in the sense that all the cases discussed in I 
correspond to tilt angle equal to zero. The
expressions for the elastic constants are derived by comparing the 
elastic free energy found from the density functional theory with that
given in Sec II. These expressions of the elastic constants involve the 
order parameters which measure the nature and amount of ordering in the 
system and direct pair correlation function (DPCF) of an effective 
isotropic liquid. In Sec.IV  we discuss the relative magnitude of these 
constants using reasonable guess for the order parameters and the 
spherical harmonic coefficients of the DPCF.

\section { Continuum Theory for Elastic Free Energy Density }

An undeformed smectic has parallel and equidistant layers and position
independent director. When some small amplitude long wavelength 
distortions are imposed on this ideal state, layers may get displaced
and curved. Let $u(x,y,z)$ represent displacements of the layer 
normal to their planes. A layer at $z_0$ before displacement is now at
\begin{equation}
{\bf z}(x,y) = {\bf z}_0 + u(x,y,z)
\end{equation}

We choose a space-fixed (SF) frame $S^0$ such that its $z$-axis is 
along the normal to the unperturbed smectic layers (see Fig.1). The 
system is described by (i) the angle $\Omega_z$ of director $\hat{\bf c}$ 
with respect to $0x$-axis and (ii) the vertical displacement $u$ of 
the layers whose derivatives 
\begin{equation}
\Omega_x = \frac{\partial u}{\partial y} \; {\rm and } \; 
\Omega_y = -\frac{\partial u }{\partial x}
\end{equation}
represent small angles of rotation about $x$ and $y$-axes, respectively.

Since the free energy of a system with broken continuous symmetry 
is invariant with respect to spatially uniform displacements and 
rotations that take the system from one point in the ground state
manifold to another, the elastic free energy must be function of 
derivatives of $\Omega$ or second order derivatives of $u$ which
correspond to the curvature of the planes. The contribution to 
free energy density due to deformation is called elastic free 
energy density $f_e(r)$.  To find  $f_e(r)$ one expands
$f(r)$ that belongs to the deformed state around the ideal state
having the free energy density $f_u$. The expansion is expressed 
in terms of the spatial derivative of the order parameter fields.  
The elastic continuum theory deals only with small spatial derivatives.
Consequently, only the lowest order terms in the expansions are
taken into account.

Because of the symmetry of Sm C the free energy density must be 
invariant of the simultaneous transformations $x \to -x, \;
y \to -y \; {\rm and } \; z \to -z$ (due to centre of symmetry)
and by the transformation $y \to -y$ as the vertical $x0z$ plane is
the plane of symmetry. Thus for the elastic free energy density
one has \cite {4}

\begin{equation}
f_e(r) = f_{dc}(r) + f_{dk}(r) + f_{dck}(r) + f_{dl}(r)   
\end{equation}
where $f_{dc}$ represents contribution arising due to
distortion of director $\hat{\bf c}$ while $f_{dk}$ is associated 
with curvature of layers. $f_{dck}$ represents the contribution 
arising due to coupling of different deformation modes whereas
$f_{dl}$ represents the contribution due to variation
in layer thickness as in Sm A. Thus \cite{1}

\begin{equation}  
f_{dl} = \frac{1}{2}\bar{B} \gamma^2
\end{equation}

where $\gamma  = \frac{\partial u}{\partial z}$ describes the variation 
in the thickness of layers.

To derive expressions for $f_{dc}, f_{dk}, f_{dck}$  we
consider director triad formed by three orthonormal unit vectors
$\hat{\bf c}, \; \hat{\bf k}, \; \hat{\bf p} $ 
and let at origin the director $\hat{\bf c}$ is along the $x$-axis of SF 
frame, $ \hat {\bf k}$ along $z$ - axis, {\it i.e.} along normal to the 
layers and $\hat{\bf p}$ along $y$-axis, {\it i.e.} 
\begin{equation}
\hat{\bf c}_0 = (1,0,0), \;\; \hat{\bf p}_0 = (0,1,0), \;\; 
\hat{\bf k}_0 = (0,0,1)
\end{equation}

The orientation of the director triad at a neighbouring point is rotated
with respect to the space-fixed coordinate frame due to deformation and 
is given as
\begin{equation}
\hat{\bf c} = (1,c_y,c_z), \;\; \hat{\bf p} = (p_x,1,p_z), \;\;
\hat{\bf k} = (k_x,k_y,1)
\end{equation}

Because of orthonormality condition of director triad
\begin{equation}
p_x = -c_y, \;\; p_z = -k_y, \;\;  c_z =- k_x 
\end{equation}

For small deviations we have 
\begin{equation}
k_x = -\frac{\partial u}{\partial y} = \Omega_y, \;\;
k_y = -\frac{\partial u}{\partial x} = - \Omega_x, \;\;
{\rm and } \;  c_y = \Omega_z 
\end{equation}

Combining the three orthonormal vectors $\hat{\bf c}, \; \hat{\bf k}, \; 
\hat{\bf p}$ we can form nine invariants. 
The elastic free energy density can be expressed
in terms of these invariants \cite{5}
\begin{eqnarray}
D_{11} & = & c_{\alpha} p_{\beta} \partial_{\alpha} k_{\beta}
= \hat{\bf c}.{\bf \nabla}\hat{\bf k}. \hat{\bf p} = -\frac{\partial 
\Omega_x}{\partial x} \nonumber \\
D_{21} & = & p_{\alpha} p_{\beta} \partial_{\alpha} k_{\beta}
= \hat{\bf p}.{\bf \nabla}\hat{\bf k}. \hat{\bf p} = -\frac{\partial 
\Omega_x}{\partial y}\nonumber \\
D_{31} & = & k_{\alpha} p_{\beta} \partial_{\alpha} k_{\beta}
= \hat{\bf k}.{\bf \nabla}\hat{\bf k}. \hat{\bf p} = -\frac{\partial 
\Omega_x}{\partial z}
= -\frac{{\partial}^2u}{\partial x \partial z} \nonumber \\
D_{12} & = & c_{\alpha} k_{\beta} \partial_{\alpha} c_{\beta}
= -\hat{\bf c}.{\bf \nabla}\hat{\bf k}.\hat{\bf c} = -\frac{\partial 
\Omega_y}{\partial x}\nonumber \\
D_{22} & = &p_{\alpha} k_{\beta} \partial_{\alpha} c_{\beta}
= -\hat{\bf p}.{\bf \nabla}\hat{\bf k}.\hat{\bf c} = -\frac{\partial 
\Omega_y}{\partial y} = D_{11}\nonumber \\
D_{32} & = & k_{\alpha} k_{\beta} \partial_{\alpha} c_{\beta}
= -\hat{\bf k}.{\bf \nabla}\hat{\bf k}.\hat{\bf c} = 
-\frac{\partial \Omega_y}{\partial z}
= -\frac{{\partial}^2u}{\partial y \partial z}\nonumber \\
D_{13} & = &c_{\alpha} c_{\beta} \partial_{\alpha} p_{\beta}
= -\hat{\bf c}.{\bf \nabla}\hat{\bf c}. \hat{\bf p} = -\frac{\partial 
\Omega_z}{\partial x} \nonumber \\
D_{23} & = &p_{\alpha} c_{\beta} \partial_{\alpha} p_{\beta}
= -\hat{\bf p}.{\bf \nabla}\hat{\bf c}.\hat{\bf p} = -\frac{\partial 
\Omega_z}{\partial y} \nonumber \\
D_{33} & = &k_{\alpha} c_{\beta} \partial_{\alpha} p_{\beta}
= -\hat{\bf k}.{\bf \nabla}\hat{\bf c}.\hat{\bf p}  = -\frac{\partial 
\Omega_z}{\partial z} 
\end{eqnarray}

Note that the invariants $D_{31}$, and $D_{32}$ correspond to 
variation of layer compression (or dilation) along  $y$ and
$x$-axes, respectively :
\begin{eqnarray}
D_{31} & = &  -\frac{\partial\Omega_x}{\partial z}
 =  -\frac{{\partial}^2u}{\partial y \partial z} 
 =  -\frac{\partial \gamma}{\partial y}
 =  -\hat{\bf p}.{\bf \nabla}{\bf \gamma}  \nonumber \\
D_{32} & = & -\frac{\partial \Omega_y}{\partial z}
 = -\frac{{\partial}^2u}{\partial x \partial z} 
 =  -\frac{\partial \gamma}{\partial x}
 =  -\hat{\bf c}.{\bf \nabla}{\bf \gamma}  
\end{eqnarray}
These terms are omitted as their contributions are small compared
to linear term $f_{dl}$ (see Eqn. (2.4)). We are therefore left with
six independent invariants.

The elastic free energy density due to orientational deformation
of Sm C can be written in terms of these six independent invariants.
Each term should, however, satisfy the symmetry elements of Sm C.
Note that $(\hat{\bf k},\; \hat{\bf c})$ plane is a symmetry plane, 
$\hat{\bf p}$ is $C_2$ axis and there is also inversion symmetry. 
We can therefore
write elastic free energy density as a sum of square of each 
invariant and their products taken in such a way that $\hat{\bf k}$ and 
$\hat{\bf c}$ enter the expansion in an even number of times ({\it i.e.} 
$\hat{\bf k}$ and $\hat{\bf c}$ are considered without discrimination) and 
$\hat{\bf p}$ which also appear even number of times but on its own. Thus
\begin{mathletters}
\begin{eqnarray}
f_{dc} & = & \frac{1}{2} B_1 D_{13}^2 + \frac{1}{2} B_2 D_{23}^2 +
         \frac{1}{2} B_3 D_{33}^2 + B_{13}
          D_{13}D_{33} \\
f_{dk} & = & \frac{1}{2}A{(D_{11} + D_{22})}^2 + \frac{1}{2}A_{12}
          D_{12}^2 + \frac{1}{2}A_{21}D_{21}^2  \\
f_{dck} & = & C_1 D_{11}D_{13} + C_2 D_{21}D_{23} + C_3 D_{12}D_{23}
        +  C_4 D_{12}D_{21}
\end{eqnarray} 
\end{mathletters}

Since 
\begin{eqnarray}
D_{12}D_{23} = \left(\frac{\partial \Omega_y}{\partial x}\right)
               \left(\frac{\partial \Omega_z}{\partial y}\right)    
= \frac{\partial}{\partial y}\left(\Omega_z\frac{\partial
 \Omega_y} {\partial x}\right) -  \frac{\partial}{\partial x}
\left(\Omega_z\frac{\partial \Omega_y} {\partial y}\right)
+ \frac{\partial \Omega_z} {\partial x}
  \frac{\partial \Omega_y} {\partial y},
\end{eqnarray}
it is the sum of two gradient terms (surface terms) and a term
analogous to $D_{11} D_{13}$. Similar decomposition can also be 
made for $D_{12}D_{21}$. We therefore have only two independent
terms in the expression of $f_{dck}$ due to coupling between
the deformation of director {\bf c} and the curvature of smectic layers,
as given in ref.[4]. 

Since $f_{dc}$ involves director $\hat{\bf c}$ and can be written
in terms of $({\bf\nabla}.\hat{\bf c})^2, (\hat{\bf c}.{\bf\nabla}\times 
\hat{\bf c})^2$ and $(\hat{\bf c}\times{\bf\nabla}\times\hat{\bf c})^2$ as
\begin{eqnarray}
({\bf\nabla}.\hat{\bf c})^2  & = &  \left({\frac{\partial \Omega_z}
{\partial y}}\right)^2 = D_{23}^2 \nonumber \\
(\hat{\bf c}.{\bf \nabla}\times\hat{\bf c})^2
 & = & \left(\frac{\partial \Omega_z} {\partial z}\right)^2  = D_{33}^2 
\nonumber \\
(\hat{\bf c}\times{\bf \nabla}\times\hat{\bf c})^2 & = & 
\left ( \frac{\partial \Omega_z} {\partial x} \right)^2 = D_{13}^2 
\nonumber \\   
{\rm and} & & \nonumber \\
(\hat{\bf k}.{\bf \nabla}\times\hat{\bf c})(\hat{\bf c}
.{\bf \nabla}\times\hat{\bf c})
 & = & \left(\frac{\partial \Omega_z} {\partial x}\right)
\left(\frac{\partial \Omega_z} {\partial z}\right)   = D_{13}D_{33}
\end{eqnarray}

Therefore the constants $B_1$, $B_2$, $B_3$, $B_{13}$ can be identified
as elastic constants associated with bend, splay, twist and coupled
twist-bend modes of deformations of director $\hat{\bf c}$ and have the
dimension of energy per unit length (dynes). The last term in Eq.(2.11a)
 is allowed because of the Sm C is invarient under the simulataneous 
transformation 
$ \hat {\bf n} \rightarrow -\hat {\bf n} $ and $\hat {\bf c}
\rightarrow -\hat {\bf c} $. These constants 
are expected to be comparable in magnitude to the Frank elastic
constants of a homologous nematic phase.

The contribution $f_{dk}$ is analogous to Sm A and is due to 
layer deformation (curvature). The term $A$, $A_{12}$ and 
$A_{21}$ represent splay contribution of a system of a monoclinic
symmetry as Sm C has $C_{2h}$ symmetry in contrast to $D_{\infty h}$
of Sm A. The value of these constants are therefore expected to 
be of the same order of magnitude as in Sm A phase.  
$\bar{B}$ (see Eq.(2.4)) which has dimension of 
energy per unit volume is associated with possible change of 
interlayer distance.

Since Sm C$^*$ has no inversion symmetry, its elastic free energy 
density may have terms linear in invariants of Eq (2.9). 
However, due to symmetry requirements only those invariants
which involve $\hat{\bf k}$ and $\hat{\bf c}$ in pairs without 
discrimination and $\bf{\hat p}$ alone contribute to linear terms. 
Thus
\begin{eqnarray}
f_d^{(1)}= - D_1(\hat{\bf c}.{\bf \nabla} \hat{\bf c}.\hat{\bf p}) + 
            D_2(\hat{\bf p}.{\bf \nabla} \hat{\bf k}.\hat{\bf c})
          - D_3(\hat{\bf k}.{\bf \nabla} \hat{\bf c}.\hat{\bf p}) 
\end{eqnarray}
where superscript (1) stands for linear terms. In terms of  
${\bf \Omega}$ vector field

\begin{eqnarray}
f_d^{(1)} =  D_1\left(\frac{\partial \Omega_z}{\partial x}\right)
         + D_2\left(\frac{\partial \Omega_x}{\partial x}\right)
         + D_3\left(\frac{\partial \Omega_z}{\partial z}\right)
\end{eqnarray}  
 
Out of the three terms in Eq.(2.15) the term $D_3$ is most 
important as it corresponds to simple twist of director
$\hat{\bf c}$. The terms $D_1$ corresponds to bend of director 
$\hat{\bf c}$ while $D_2$ tends to transform a flat layer into 
a twisted ribbon. The terms $D_1$ and $D_2$ are expected to be 
negligible in comparison to $D_3$ term.

If $q_0 = -\left(\frac{D_3}{B_3}\right) $ where $B_3$ is defined in 
Eq.(2.11a) the contribution $f_{dc}(r)$ for Sm C$^*$ is modified to
\begin{eqnarray}
f_{dc}(r) = \frac{1}{2} B_1\left(\frac{\partial \Omega_z}
            {\partial x}\right)^2 + \frac{1}{2} B_2\left(\frac
            {\partial \Omega_z}{\partial y}\right)^2
         + \frac{1}{2} B_3 \left(\frac{\partial \Omega_z}
            {\partial z} + q_0\right)^2
          +B_{13}\left(\frac{\partial \Omega_z}
            {\partial x }\right )\left(\frac{\partial \Omega_z}
            {\partial z }\right)    
\end{eqnarray} 
while other terms of Eq.(2.11) remain same. The ground state
of Sm C$^*$ has a helical structure. The $B_3$ term can therefore
be considered to be a contribution due to deformation from this
uniform helical structure.

The expressions for the elastic constants are obtained by comparing
the expression of elastic free energy density obtained from the 
continuum theory and from a microscopic theory involving molecular
parameters. In order to make this comparison possible we have to 
rewrite Eq.(2.16) in terms of distortion vectors defined below.

Since in the ground state the directors are position independent
we take $S^0$ to be the unperturbed director frame at a point $R$.
Let $S^0$ be described by the unit vectors ${\bf N}_0^{(j)}$ such that
(see Eq. (2.5))

\begin{eqnarray}  
{\bf N}_0^{(1)} = \hat{\bf c}_0, \;\;\; {\bf N}_0^{(2)}= \hat{\bf p}_0, 
\;\;\; {\bf N}_0^{(3)}= \hat{\bf k}_0, \;\;\;   
\end{eqnarray}

Let S be the perturbed director frame described by the unit vector
${\bf N}^{(j)}$ such that (see Eq. (2.6))

\begin{eqnarray}
{\bf N}^{(1)} = \hat{\bf c} = \hat{\bf c}_0 + {\bf\sigma}^{(1)},\;\;
{\bf N}^{(2)} = \hat{\bf p} = \hat{\bf p}_0 + {\bf\sigma}^{(2)},\;\;
{\bf N}^{(3)} = \hat{\bf k} = \hat{\bf k}_0 + {\bf\sigma}^{(3)},\;\;  
\end{eqnarray}  
                    
where ${\bf \sigma}^{(j)}$ may be regarded as the distortion vectors, and let
\begin{equation}
{\bf N}_{k}^{(j)} = {\bf N}^{(j)}.{\bf N}_0^{(k)} = \delta_{jk} + 
{\bf \sigma}_{k}^{(j)}
\end{equation}
be the components of the perturbed directors with respect to $S^0$
frame.

The transformation from $S^0$ to $S$ is effected by $3\times3$ 
orthogonal matrix $T$ with element
\begin{displaymath}
T_{jk} = {\bf N}_{k}^{(j)} = \delta_{jk} + \sigma_{k}^{(j)}
\end{displaymath}
such that 
\begin{displaymath}
{\bf N}^{(j)} = T_{jk} {\bf N}_0^{(k)} = {\bf N}_0^{(j)} + 
{\bf \sigma}_{k}^{(j)} {\bf N}_0^{(k)}
\end{displaymath}
The orthogonality condition $T_{ij} T_{ik} = \delta_{jk}$ leads to
\begin{eqnarray}
\sigma_j^{(k)} & = & - \sigma_k^{(j)} - \sigma_j^{(i)} 
\sigma_k^{(i)} \; \; \; \; {\rm for} \; \; j\ne k \nonumber \\
\sigma_j^{(j)} & = & - \frac{1}{2} {\sum_{i}}^{\prime} {\sigma_j^{(i)}}^2  
\; \; \; \;\; \; \; \; \; \; {\rm for} \;  \; j = k 
\end{eqnarray}  
where prime on summation indicates no summation over $j$. Thus for 
small deformations 
\begin{displaymath}
\sigma_j^{(k)} \sim O \; {\rm and} \; \sigma_j^{(j)} \sim O^2
\end{displaymath}
where $O$ means first order small quantity.

From above relations it follows that $\sigma_2^{(1)}$, $\sigma_1^{(3)}$ 
and $\sigma_3^{(2)}$ can be regarded as basic independent distortion 
variables; all as a function of $R$. Every other component can be 
constructed with the help of Eq.(2.20). 

From above discussions we find 
\begin{eqnarray}
c_y  & = & \Omega_z = \sigma_2^{(1)} \nonumber \\
k_x  & = & \Omega_y = \sigma_1^{(3)} \nonumber \\
k_y  & = & - \Omega_x = -\sigma_3^{(2)} 
\end{eqnarray}  
Using these relations we rewrite Eq. (2.11) as 
\begin{mathletters}
\begin{eqnarray}
F_{dc} & = & \int d{\bf R} \left [ \frac{1}{2} B_1 \left (
\frac{\partial\sigma_2^{(1)}}{\partial x}\right )^2 
+ \frac{1}{2} B_2 \left (\frac{\partial\sigma_2^{(1)}}{\partial y}\right)^2 
\nonumber \right. \\
& & + \left.\frac{1}{2} B_3 \left(\frac{\partial\sigma_2^{(1)}}
{\partial z}\right)^2 + B_{13} \left(\frac{\partial
\sigma_2^{(1)}}{\partial x}\right) \left (\frac{\partial
\sigma_2^{(1)}}{\partial z}\right)\right] \\
F_{dk} & = & \int d{\bf R} \left [ \frac{1}{2} A \left (\frac{\partial
\sigma_3^{(2)}}{\partial x}\right)^2 + 
\frac{1}{2} A_{12} \left (\frac{\partial\sigma_1^{(3)}} 
{\partial x}\right)^2 + \frac{1}{2} A_{21} \left (\frac{\partial\sigma_3^{(2)}} 
{\partial y}\right)^2 \right] \\
F_{dck} & = & \int d{\bf R} \left [ C_1 \left(\frac{\partial\sigma_3^{(2)}}
{\partial x}\right)\left(\frac{\partial\sigma_2^{(1)}}{\partial x}\right) +
C_2 \left(\frac{\partial\sigma_3^{(2)}}{\partial y}\right)
\left (\frac{\partial \sigma_2^{(1)}}{\partial y}\right)\right]
\end{eqnarray}
\end{mathletters}
It is often convenient to assume long wavelength deformations 
in space fixed frame having the form 
\begin{equation}
\sigma_{k}^{(j)} = B_{k}^{(j)} \sin({\bf q}.{\bf R})
\end{equation}
where $B_k^{(j)}$ are small amplitudes and $q$ is the wave vector.
Substituting this into Eq. (2.22) leads to 
\begin{mathletters}
\begin{eqnarray}
\frac{F_{dc}}{V} & = & \frac{1}{4} \left[B_1 (B_2^{(1)})^2 q_x^2 + 
B_2 (B_2^{(1)})^2 q_y^2 + B_3 (B_2^{(1)})^2 q_z^2 + 
B_{13} (B_2^{(1)})^2 q_x q_z\right] \\
\frac{F_{dk}}{V} & = & \frac{1}{4} \left [A (B_3^{(2)})^2 q_x^2 + A_{12} 
(B_1^{(3)})^2 q_x^2 + A_{21} (B_3^{(2)})^2 q_y^2\right] \\
\frac{F_{dck}}{V} & = & \frac{1}{2} C_1 (B_3^{(2)} B_2^{(1)}) q_x^2 +
\frac{1}{2} C_2 (B_3^{(2)} B_2^{(1)}) q_y^2 
\end{eqnarray} 
\end{mathletters}
In writing above equations we have used the relations 
\begin{displaymath}
\int d{\bf R} \sin^2({\bf q}.{\b R}) = \int d{\bf R} 
\cos^2({\bf q}.{\bf R}) = \frac{V}{2}
\end{displaymath}
and
\begin{displaymath}
\int d{\bf R} \sin({\bf q}.{\bf R}) \cos({\bf q}.{\bf R}) = 0
\end{displaymath}

\section{Density-functional approach}
\subsection{Expression for the distortion free energy}

A liquid crystalline phase is characterized by the spatial and 
orientational configuration of molecules. At the phase transition 
point these configurations undergo a modification, {\it i.e.}, 
abrupt change may take place in the symmetries of the system. The
molecular configurations of most ordered phases are adequately 
described by the single particle density distribution $\rho ({\bf x})$.
The vector ${\bf x}$ is taken  here to indicate both the location 
${\bf r}$ of the centre of a molecule and its relative orientation ${\bf 
\Omega}$ described by Euler angles ${\bf \phi}$, $\theta$, $\psi$. 
For an isotropic uniform system $\rho({\bf x})$ is independent of 
positions and orientations. 

The single particle density distribution $\rho({\bf x})$ provides 
us with a convenient variational quantity to specify an arbitrary 
state of a system. One may consider a variational thermodynamic
potential $W (T,P, [\rho({\bf x})])$ as a function of $\rho({\bf x})$.
The equilibrium state of the system at a given $T$ and $P$ is 
described by the density  $\rho(T,P,{\bf x})$ corresponding to 
the minimum of $W$ with respect to $\rho({\bf x})$. This forms 
the basis of the density-functional theory \cite{2}. 

The basic thermodynamic potential used to determine the isothermal 
elastic properties of a system consisting of $N$ particles in 
volume $V$ at temperature $T$, is the Helmholtz free energy $A[\rho]$.
Elasticity is associated with the behaviour of $A[\rho]$ with respect
to small deformation of the system away from its equilibrium (ground)
state \cite{1}.

In the density-functional formalism the free energy of a system is 
expressed in terms of the direct correlation function of the 
medium \cite{2}. 
\begin{equation}
\beta A[\rho] = \beta A_{id}[\rho] + \beta \Delta A[\rho]
\end{equation}
where the ideal gas part 
\begin{equation}
\beta A_{id}[\rho] = \int d{\bf x} \; \rho({\bf x}) \; \{ \ln(\rho({\bf x})
\Lambda) -1 \} 
\end{equation}                                                               
and $\beta \Delta A[\rho]$ is the excess reduced free energy arising 
due to intermolecular interactions. Here $\beta = (k_B T)^{-1}$ where 
$k_B$ is the Boltzmann constant and $T$ is the temperature. 
In the weighted density-functional approach 
\begin{equation}
\beta \Delta A[\rho] = -\frac{1}{2} \int d{\bf x_1} \int d{\bf x_2}
 \; \rho({\bf x_1}) c({\bf x_1},{\bf x_2}, \bar{\rho}) \rho({\bf x_2}) 
\end{equation}                                                                  
Here the function $c$ represents the direct pair correlation function
(DPCF) of an effective isotropic fluid. The effective density $\bar{\rho}$
is found using the relation \cite{6}
\begin{equation}
\overline{\rho}[\rho] = \frac{1}{\rho_0 V} \int d{\bf x_1} \int d{\bf x_2}
\; \rho({\bf x_1}) \; \rho({\bf x_2}) \; \omega({\bf x_1},{\bf x_2};
\overline{\rho}) 
\end{equation}
where $\rho_0$ is the averaged density of the ordered phase and
$\omega$ is a weight factor. $\overline{\rho}[\rho]$ is viewed here as
a functional of $\rho({\bf x})$. To ensure that the approximation becomes
exact in the limit of a uniform system, the weight factor $\omega$
must satisfy the normalization condition
\begin{displaymath}
\int d{\bf x_2} \; \omega({\bf x_1},{\bf x_2};\overline{\rho}) = 1 
\end{displaymath}
Requiring that $\omega$ must satisfy
\begin{equation}
-c^{(2)}({\bf x_1},{\bf x_2};\rho_0) = \lim_{\rho\rightarrow\rho_0}
\frac{\delta^2(\beta \Delta A)}{\delta\rho({\bf x_1}) {\delta}\rho({\bf x_2})}
\end{equation}
exactly, one finds                                                              
\begin{equation}
\omega({\bf x_1},{\bf x_2};\overline{\rho}) = -\frac{1}
{2 \Delta a^{\prime}(\overline{\rho})} \left [ \beta^{-1} \;
c({\bf x_1},{\bf x_2};\overline{\rho}) + \frac{1}{V} \overline{\rho}
{\Delta}a^{\prime\prime}(\overline{\rho}) \right ] 
\end{equation}
where $\Delta a(\overline{\rho})$ is the excess free energy per particle
and primes on it denote derivatives with respect to density.                

The contribution to the free energy due to deformation is given as \cite{1}
\begin{eqnarray}
\beta \Delta A_e[\rho]  & = & \beta (\Delta A_d[\rho] - 
\Delta A_u[\rho]) \nonumber \\
& = & -\frac{1}{2} \int d{\bf x_1} \int d{\bf x_2}
\; \left[\rho_d({\bf x_1}) \rho_d({\bf x_2}) - \rho({\bf x_1}) 
\rho({\bf x_2})\right] \; c({\bf x_1},{\bf x_2}; \overline{\rho}) .
\end{eqnarray}                                                                  
where $\beta \Delta A_u[\rho]$ and $\rho({\bf x})$ are, respectively,
the excess free energy and single particle density distribution 
of the ground state of the ordered phase. The subscript $d$ refers 
to the corresponding quantities of the deformed state. In writing 
Eq.(3.7) it is assumed that the direct pair correlation functions 
do not change due to deformation. 

Since the isotropic fluid DPCF is an invariant pairwise function 
it has an expansion in SF frame of the form 
\begin{eqnarray}
c({\bf r}, {\bf \Omega}_1, {\bf \Omega}_2)  =  
\sum_{l_1 l_2 l} \;  \sum_{m_1 m_2 m} & & 
c(l_1,l_2,l;n_1,n_2;r)\; C_g(l_1,l_2,l;m_1,m_2,m) \nonumber \\
& & D_{m_1,n_1}^{l_1 *} (\Omega_1)  D_{m_2,n_2}^{l_2 *}(\Omega_2) 
Y_{lm}^{*}(\hat{r}) 
\end{eqnarray}
where $C_g(l_1,l_2,l;m_1,m_2,m)$ are the Clebsch-Gordon coefficients,
$c(l_1,l_2,l;n_1,n_2;r)$ the harmonic expansion coefficient of the 
DPCF. $\hat{\bf r} = {\bf r}/\mid {\bf r} \mid$, is a unit vector 
along the intermolecular axis. 

The elastic constants are defined by the second-order term of the 
expansion of the free energy of the deformed state around the 
free energy of the equilibrium (ground) state in the ascending powers 
of a parameter, which measures the deformation. The first term of this
expansion is balanced by the equilibrium stresses of the ground 
state. One defines the elastic free energy per unit volume as 
\begin{displaymath}
\frac{E_e}{V} = \frac{1}{V} [ \Delta A_e[\rho] + P(V_d - V)]
\end{displaymath}
where $V_d$ is the volume of the deformed sample and $P$ the 
isotropic pressure. 

For an ordered (liquid crystalline) phase $\rho ({\bf x})$ is 
expressed in terms of the Fourier series and the Wigner rotation 
matrices \cite{2} as 

\begin{mathletters}
\begin{eqnarray}
\rho ({\bf x}) & = & \rho({\bf r},{\bf \Omega}^{\prime}) = \rho_0 \sum_{G}
\sum_{lmn} Q_{lmn}({\bf G}) \;\exp[i{\bf G}.{\bf r}] \;
D_{mn}^l({\bf \Omega}^{\prime}) \\
{\rm where \; the \; expansion \; coefficients} & & \nonumber \\
Q_{lmn}({\bf G}) & = & \frac{2l+1}{N} \int d{\bf r} \int d{\bf \Omega^{\prime}}
\rho ({\bf r},{\bf \Omega^{\prime}}) \exp(-i {\bf G}.{\bf r}) 
D_{mn}^{l*}({\bf \Omega^{\prime}}) 
\end{eqnarray}
\end{mathletters}
Here $\{{\bf G}\}$ are the reciprocal lattice vectors (RLV's) of the 
positionally 
ordered structure that might be present in the system and $\rho_0$ 
is the mean number density of the system. All the angles which
appear in Eq.(3.9) refer to a coordinate frame the $z$-axis of which 
is along the director $\hat{\bf n}$. Using the rotational property 
of the $D$ matrices \cite{7} we rewrite Eq.(3.9a) so that all the 
angles appearing in it refer to space fixed coordinate frame 
shown in Fig.1. Thus 
\begin{equation}
\rho({\bf r},{\bf \Omega}) = \rho_0 \sum_{G} \sum_{lmn} \sum_q  Q_{lmn}(G) 
\exp[i{\bf G}.{\bf r}] D_{qm}^{l*}(\hat{\bf n}) D_{qn}^{l} ({\bf \Omega}) \\
\end{equation}

In the limit of long wavelength distortion the magnitude of the 
order parameters are assumed to remain unchanged. The changes 
occur in the direction of the directors making them position 
dependent and in the RLV's ${\bf G}$. The RLV's ${\bf G_d}$ of the strained 
structure are related to ${\bf G}$ of the unstrained structure as 
\begin{equation}
{\bf G}_d = ({\bf I} + {\rm \large \bf \epsilon})^{-1} {\bf G} 
\end{equation}
here ${\rm \large \bf \epsilon }$ is a strain matrix which governs 
the change in position. Thus for a deformed state
\begin{equation}
\rho_d({\bf r},{\bf\Omega}) = \rho_0 \sum_{G} \sum_{lmn} \sum_q 
 Q_{lmn}({\bf G}) \exp(i {\bf G}_d.{\bf r}) D_{qm}^{l*} (\hat{\bf n}_d) 
D_{qn}^{l} ({\bf \Omega})
\end{equation}                               
where $\hat{\bf n}_d (r)$ indicates the direction of director at 
$r$ of the deformed state. In the deformed state the directors
orientations become position dependent. One uses the rotational
properties of $D$ matrices to re write Eq.(3.11) in terms of the 
director of a chosen point. Substituting Eqs.(3.8)-(3.12)
into Eq.(3.7) and after some simplifications we get 
\begin{eqnarray}
\frac{\beta \Delta A_d}{V} & = & -\frac{1}{2} \rho_0^2 \sum_{l_1 l_2 l}
\sum_{m_1^{\prime} m_2^{\prime} m^{\prime}} \sum_{m \lambda} 
\sum_{n_1 n_2} \sum_G \frac{Q_{l_1 m_1^{\prime} n_1}(G)\; 
Q_{l_2 m_2^{\prime} n_2}(-G)}{(2l_1+1) (2l_2+1)}\nonumber \\
& & C_g(l_1 l_2 l; m_1^{\prime} m^{\prime} \lambda) \;
D_{m \lambda}^{l*}(\hat{\bf n}) \int d{\bf r}\; [\exp(i{\bf G}_d.{\bf r}) 
D_{m^{\prime} m_2^{\prime}}^{l_2}(\Delta \chi) \nonumber \\ 
& - & \delta_{m^{\prime} m_2^{\prime}} \exp(i {\bf G}.{\bf r})] 
\;c(l_1 l_2 l; n_1 n_2; r) Y_{lm}^{*}(\hat{{\bf r}})
\end{eqnarray}
where $\hat{{\bf n}}$ indicates the orientation of the director 
at the reference point ${\bf R}$ and $\Delta \chi$ the angle 
between the directors at ${\bf R}$ and ${\bf R}+{\bf r}$. Eq.(3.13)
presents a general expression for the distortion free energy 
density for the continuous symmetry broken phases with tilted
orientational order in the limit of long wavelength distortion. 
Since for uniaxial phases the director is along the $z$-axis of the 
SF coordinate frame, $D_{m \lambda}^{l*}(n=0) = \delta_{m \lambda}$.
When this is substituted in Eq.(3.13) it reduces to Eq.(3.9) of I.

Eq.(3.13) can be used to derive expressions for the contributions
arising due to curvature in the director orientation and dilation 
in layer thickness. The term which represents the coupling of these
two terms of distortions is neglected. Thus
\begin{equation}
\frac{\beta \Delta A_d}{V} = f_1 + f_2
\end{equation}
where 
\begin{eqnarray}
f_1 & = & -\frac{1}{2} \rho_0^2 \sum_{l_1 l_2 l}
\sum_{m_1^{\prime} m_2^{\prime} m^{\prime}} \sum_{m \lambda} 
\sum_{n_1 n_2} \sum_G \frac{Q_{l_1 m_1^{\prime} n_1}(G)\; 
Q_{l_2 m_2^{\prime} n_2}(-G)}{(2l_1+1) (2l_2+1)}
C_g(l_1 l_2 l; m_1^{\prime} m^{\prime} \lambda)\nonumber \\ 
& & D_{m \lambda}^{l*}(\hat{\bf n}) \int d{\bf r} \exp(i{\bf G}.{\bf r}) 
[D_{m^{\prime} m_2^{\prime}}^{l_2 *} (\Delta \chi) - 
\delta_{m^{\prime} m_2^{\prime}}]
Y_{lm}^{*}(\hat{{\bf r}}) \;c(l_1 l_2 l; n_1 n_2; r)
\end{eqnarray}
\begin{eqnarray}
f_2 & = & -\frac{1}{2} \rho_0^2 \sum_{l_1 l_2 l}
\sum_{m_1^{\prime} m_2^{\prime} m^{\prime}} \sum_{m \lambda} 
\sum_{n_1 n_2} \sum_G \frac{Q_{l_1 m_1^{\prime} n_1}(G)\; 
Q_{l_2 m_2^{\prime} n_2}(-G)}{(2l_1+1) (2l_2+1)}
C_g(l_1 l_2 l; m_1^{\prime} m^{\prime} \lambda)\nonumber \\ 
& & D_{m \lambda}^{l*}(\hat{\bf n}) \int d{\bf r} [\exp(i{\bf G}_d.{\bf r}) 
- \exp(i{\bf G}.{\bf r})] \delta_{m^{\prime} m_2^{\prime}}
Y_{lm}^{*}(\hat{{\bf r}}) \;c(l_1 l_2 l; n_1 n_2; r).
\end{eqnarray}

While $f_1$ includes all contributions arising due to curvature 
in directors orientations, $f_2$ the contribution due to dilation
(compression) in layer thickness. 

\subsection{Expressions for the elastic constants associated with 
directors orientations}

All the elastic constants associated with curvature in director 
orientations can be derived from the expression of $f_1$ given in 
Eq.(3.15). Using the procedure outlined in II we rewrite Eq.(3.15)
in terms of the director components. Thus 
\begin{eqnarray}
f_1 & = & -\frac{1}{2} \rho_0^2 \sum_{l_1 l_2 l}
\sum_{m_1^{\prime} m_2^{\prime} m^{\prime}} \sum_{m \lambda} 
\sum_{n_1 n_2} \sum_G \frac{Q_{l_1 m_1^{\prime} n_1}(G)\; 
Q_{l_2 m_2^{\prime} n_2}(-G)}{(2l_1+1) (2l_2+1)} 
D_{m \lambda}^{l*}(\hat{\bf n})  \int dr \;  r^4 \int d\hat{\bf r} \nonumber \\
& & \left [\frac{1}{2} \mid B_1^{(3)}\mid^2 \{ (a_{11} + a_{12}) 
(q_z^2 - q_x^2) (\hat{\bf r}.\hat{\bf x}) (\hat{\bf r}.\hat{\bf z})
+ (a_{23} + a_{21} + a_{22}) (q_x^2 (\hat{\bf r}.\hat{\bf x})^2 +
\cdot \cdot \cdot \cdot)\}\right. \nonumber \\
& + &\frac{1}{2} \mid B_2^{(3)}\mid^2 \{ i (a_{12} - a_{11}) 
q_y^2 (\hat{\bf r}.\hat{\bf y}) (\hat{\bf r}.\hat{\bf z}) 
+ (a_{23} - a_{21} - a_{22}) 
(q_x^2 (\hat{\bf r}.\hat{\bf x})^2 +q_y^2 (\hat{\bf r}.\hat{\bf y})^2 +
\cdot \cdot \cdot \cdot)\} \nonumber \\
& + &\frac{1}{2} \mid B_2^{(1)}\mid^2 \{ i b_2 (q_x^2 - q_y^2) 
(\hat{\bf r}.\hat{\bf x}) (\hat{\bf r}.\hat{\bf y}) - \frac{1}{2} b_2^2 
(q_x^2 (\hat{\bf r}.\hat{\bf x})^2 + q_y^2 (\hat{\bf r}.\hat{\bf y})^2 +
q_z^2 (\hat{\bf r}.\hat{\bf z})^2) \nonumber \\
& - & b_2^2 q_z q_x (\hat{\bf r}.\hat{\bf z}) 
(\hat{\bf r}.\hat{\bf x}) + \cdot \cdot \cdot  \cdot\}     
+ B_2^{(1)} B_2^{(3)} \{ -\frac{1}{4} (a_{11} + a_{12}) 
(q_x^2 (\hat{\bf r}.\hat{\bf x})^2 + q_y^2 (\hat{\bf r}.\hat{\bf y})^2 + 
\cdot \cdot \cdot \cdot ) \nonumber \\
& + & \frac{i}{2} (a_{11} - a_{12}) ((q_x^2 - q_y^2) (\hat{\bf r}.\hat{\bf x}) 
(\hat{\bf r}.\hat{\bf y}) + \cdot \cdot \cdot \cdot) 
+ \frac{i b_2}{2} ((q_z^2 - q_y^2) 
(\hat{\bf r}.\hat{\bf y}) (\hat{\bf r}.\hat{\bf z}) + 
\cdot \cdot \cdot \cdot)  \nonumber \\
& - & \left .\frac{1}{2} m_2^{\prime} (a_{11} - a_{12}) 
(q_x^2 (\hat{\bf r}.\hat{\bf x})^2 +q_y^2 (\hat{\bf r}.\hat{\bf y})^2 +
\cdot\cdot \cdot \cdot )\} + \cdot \cdot \cdot \cdot\right] 
\exp(i{\bf G}.{\bf r}) Y_{lm}^{*}(\hat{{\bf r}}) \; 
c(l_1 l_2 l; n_1 n_2; r)
\end{eqnarray}
where
\begin{eqnarray*}
a_{11} & = & -\frac{1}{2} \{(l_2 + m_2^{\prime} + 1) (l_2 - m_2^{\prime})
\}^{\frac{1}{2}} C_g(l_1 l_2 l; m_1^{\prime}, m_2^{\prime}+1, \lambda) \\
a_{12} & = & \frac{1}{2} \{(l_2 - m_2^{\prime} + 1) (l_2 + m_2^{\prime})
\}^{\frac{1}{2}} C_g(l_1 l_2 l; m_1^{\prime}, m_2^{\prime}-1, \lambda) \\
a_{21} & = & \frac{1}{8} \{(l_2 + m_2^{\prime} + 2) (l_2 - m_2^{\prime}
- 1) (l_2 + m_2^{\prime} + 1) (l_2 - m_2^{\prime}) \}^{\frac{1}{2}} 
C_g(l_1 l_2 l; m_1^{\prime}, m_2^{\prime}+2, \lambda) \\
a_{22} & = & \frac{1}{8} \{(l_2 - m_2^{\prime} + 2) (l_2 + m_2^{\prime}
- 1) (l_2 - m_2^{\prime} + 1) (l_2 + m_2^{\prime}) \}^{\frac{1}{2}} 
C_g(l_1 l_2 l; m_1^{\prime}, m_2^{\prime}-2, \lambda) \\
a_{23} & = & - \frac{1}{8} \{(l_2 + m_2^{\prime}) (l_2 - m_2^{\prime} + 1) 
+ (l_2 - m_2^{\prime}) (l_2 + m_2^{\prime} +1) \} 
C_g(l_1 l_2 l; m_1^{\prime}, m_2^{\prime}, \lambda) \\
b_{21} & = & m_2^{\prime}  
C_g(l_1 l_2 l; m_1^{\prime}, m_2^{\prime}, \lambda) \\
b_{22} & = & m_2^{\prime 2}  
C_g(l_1 l_2 l; m_1^{\prime}, m_2^{\prime}, \lambda) 
\end{eqnarray*}

Since the plane wave $\exp(i{\bf G.r})$ travelling in the $z$-direction
(because of layered structure of smectic phases) is symmetrical about the
$z$-axis and can be expanded as a series of Legendre polynomial referred to 
this axis,

\begin{eqnarray}
e^{i{\bf G} z_{12}} = \sum_{l'} (i)^{l'} (2l' + 1) 
j_{l'} (Gr) P_{l'}(\cos\theta)
\end{eqnarray}

where $j_{l'}(Gr)$ are the spherical Bessel functions and $\theta$ is 
angle between the $z$-axis and intermolecular axis $r$.

After performing the angular integration over $\hat{\bf r}$ we compare 
the resulting expression with Eq.(2.24) and obtain the following 
expressions for the elastic constants.

\begin{eqnarray}
\beta B_1 & = & -\rho_0^2 \sqrt{4 \pi} \sum_{l_1 l_2 l l'}
\sum_{m_1^{\prime} m_2^{\prime}} \sum_{m \lambda} \sum_{n_1 n_2} \sum_G 
(i)^{l'} \frac{\sqrt{(2l' + 1)}}{(2l_1+1)(2l_2+1)} Q_{l_1 {m_1}^{'}n_1}(G) 
Q_{l_2 m_2^{'} n_2}(-G)  \nonumber \\
& & D_{m\lambda}^{l*}(\hat{\bf n}) J_{l_1 l_2 l l'}^{n_1 n_2} 
\left [\frac{b_{21}}{\sqrt{6}} \left(\frac{2l'+1}{2l+1}\right)^{1/2} 
C_g(2 l' l; 0 0 0) \{ C_g(2 l' l; 2 0 2) \delta_{m 2} - 
C_g(2 l' l; \underline{2} 0\underline{2}) \delta_{m \underline{2}} \} 
\right .\nonumber \\
& - & \frac{b_{22}}{6} \{ \delta_{ll'} \delta_{m0} - 
\left (\frac{2l'+1}{2l+1}\right)^{1/2} C_g(2 l' l; 0 0 0)^2 
\delta_{m0} + \sqrt{\frac{3}{2}} \left(\frac{2l'+1}{2l+1}\right)^{1/2} 
C_g(2 l' l; 0 0 0) \nonumber \\
& & \left.(C_g(2 l' l; 2 0 2) \delta_{m2} + C_g(2 l' l; \underline{2} 0 
\underline{2}) \delta_{m \underline{2}})\}\right]
\end{eqnarray}

\begin{eqnarray}
\beta B_2 & = & \rho_0^2 \sqrt{4 \pi} \sum_{l_1 l_2 l l'}
\sum_{m_1^{\prime} m_2^{\prime}} \sum_{m \lambda} \sum_{n_1 n_2} \sum_G 
(i)^{l'} \frac{\sqrt{(2l' + 1)}}{(2l_1+1)(2l_2+1)} Q_{l_1 m_1^{'}n_1}(G) 
Q_{l_2 m_2^{'} n_2}(-G)  \nonumber \\
& & D_{m\lambda}^{l*}(\hat{\bf n}) J_{l_1 l_2 l l'}^{n_1 n_2} 
\left[\frac{b_{21}}{\sqrt{6}} \left(\frac{2l'+1}{2l+1}\right)^{1/2} 
C_g(2 l' l; 0 0 0) \{C_g(2 l' l; 2 0 2) \delta_{m 2} - 
C_g(2 l' l; \underline{2} 0\underline{2}) 
\delta_{m \underline{2}} \} \right.\nonumber \\
& + & \frac{b_{22}}{6} \{ \delta_{ll'} \delta_{m0} - 
\left(\frac{2l'+1}{2l+1}\right)^{1/2} C_g(2 l' l; 0 0 0) \delta_{m0} - 
\sqrt{\frac{3}{2}} \left(\frac{2l'+1}{2l+1}\right)^{1/2} 
C_g(2 l' l; 0 0 0) \nonumber \\
& &\left. (C_g(2 l' l; 2 0 2) \delta_{m2} +
C_g(2 l' l; \underline{2} 0 \underline{2}) 
\delta_{m \underline{2}})\}\right]
\end{eqnarray}

\begin{eqnarray}
\beta B_3 & = & \rho_0^2 \sqrt{4 \pi} \sum_{l_1 l_2 l l'}
\sum_{m_1^{\prime} m_2^{\prime}} \sum_{m \lambda} \sum_{n_1 n_2} \sum_G 
\frac{(i)^{l'}}{6} \frac{\sqrt{(2l' + 1)}}{(2l_1+1)(2l_2+1)} 
Q_{l_1 m_1^{'}n_1}(G) Q_{l_2 m_2^{'} n_2}(-G) \nonumber \\
& & D_{m\lambda}^{l*}(\hat{\bf n}) J_{l_1 l_2 l l'}^{n_1 n_2} 
b_{22} \left [\delta_{ll'} \delta_{m0} + 2\left(\frac{2l'+1}{2l+1}\right)^{1/2} 
C_g(2 l' l; 0 0 0)^2 \delta_{m0}\right] 
\end{eqnarray}

\begin{eqnarray}
\beta B_{13} & = & -\rho_0^2 \sqrt{4 \pi} \sum_{l_1 l_2 l l'}
\sum_{m_1^{\prime} m_2^{\prime}} \sum_{m \lambda} \sum_{n_1 n_2} \sum_G 
(i)^{l'} \frac{\sqrt{(2l' + 1)}}{(2l_1+1)(2l_2+1)} \frac{1}{2\sqrt{6}}
\frac{1}{\sqrt{(2l+1)}} Q_{l_1 m_1^{'}n_1}(G) \nonumber \\ 
& & Q_{l_2 m_2^{'} n_2}(-G)\; D_{m\lambda}^{l*}(\hat{\bf n}) 
J_{l_1 l_2 l l'}^{n_1 n_2} b_{22}
C_g(2 l' l; 0 0 0) [C_g(2 l' l; 1 0 1) \delta_{m 1} + 
C_g(2 l' l; \underline{1} 0\underline{1}) \delta_{m \underline{1}}]
\end{eqnarray}

\begin{eqnarray}
\beta A & = & -\rho_0^2 \sqrt{4 \pi} \sum_{l_1 l_2 l l'}
\sum_{m_1^{\prime} m_2^{\prime}} \sum_{m \lambda} \sum_{n_1 n_2} \sum_G 
(i)^{l'} \frac{\sqrt{(2l' + 1)}}{(2l_1+1)(2l_2+1)} \frac{1}{3}
(a_{23} - a_{21} - a_{22}) Q_{l_1 m_1^{'}n_1}(G) \nonumber \\
& & Q_{l_2 m_2^{'} n_2}(-G)\;
D_{m\lambda}^{l*}(\hat{\bf n}) J_{l_1 l_2 l l'}^{n_1 n_2} 
\left[\delta_{ll'} \delta_{m0} -\left(\frac{2l'+1}{2l+1}\right)^{1/2} 
C_g(2 l' l; 0 0 0)^2 \delta_{m0}  + \sqrt{\frac{3}{2}}
\left(\frac{2l'+1}{2l+1}\right)^{1/2} \right. \nonumber \\
& & C_g(2 l' l; 0 0 0) \left .\{C_g(2 l' l; 2 0 2) \delta_{m 2} + 
C_g(2 l' l; \underline{2} 0\underline{2}) \delta_{m \underline{2}} \}\right] 
\end{eqnarray}

\begin{eqnarray}
\beta A_{12} & = & -\rho_0^2 \sqrt{4 \pi} \sum_{l_1 l_2 l l'}
\sum_{m_1^{\prime} m_2^{\prime}} \sum_{m \lambda} \sum_{n_1 n_2} \sum_G 
(i)^{l'} \frac{\sqrt{(2l' + 1)}}{(2l_1+1)(2l_2+1)} Q_{l_1 m_1^{'}n_1}(G) 
Q_{l_2 m_2^{'} n_2}(-G) \nonumber \\
& & D_{m\lambda}^{l*}(\hat{\bf n}) J_{l_1 l_2 l l'}^{n_1 n_2} 
[\frac{1}{\sqrt{6}} \left(\frac{2l'+1}{2l+1}\right)^{1/2} 
(a_{11} + a_{12}) C_g(2 l' l; 0 0 0) \{ C_g(2 l' l; 1 0 1) \delta_{m 1} + 
C_g(2 l' l; \underline{1} 0\underline{1}) \delta_{m \underline{1}} \} 
\nonumber \\
& + & \frac{1}{3} (a_{23} + a_{21} + a_{22}) \{ \delta_{ll'} \delta_{m0} - 
\left(\frac{2l'+1}{2l+1}\right)^{1/2} C_g(2 l' l; 0 0 0)^2 \delta_{m0} 
+ \sqrt{\frac{3}{2}} \left(\frac{2l'+1}{2l+1}\right)^{1/2} 
C_g(2 l' l; 0 0 0) \nonumber \\
& & (C_g(2 l' l; 2 0 2) \delta_{m2} +
C_g(2 l' l; \underline{2} 0 \underline{2}) \delta_{m \underline{2}})\}]
\end{eqnarray}

\begin{eqnarray}
\beta A_{21} & = & -\rho_0^2 \sqrt{4 \pi} \sum_{l_1 l_2 l l'}
\sum_{m_1^{\prime} m_2^{\prime}} \sum_{m \lambda} \sum_{n_1 n_2} \sum_G 
(i)^{l'} \frac{\sqrt{(2l' + 1)}}{(2l_1+1)(2l_2+1)} Q_{l_1 m_1^{'}n_1}(G) 
Q_{l_2 m_2^{'} n_2}(-G)  \nonumber \\
& &D_{m\lambda}^{l*}(\hat{\bf n}) J_{l_1 l_2 l l'}^{n_1 n_2} 
[-\frac{1}{\sqrt{6}} \left(\frac{2l'+1}{2l+1}\right)^{1/2} (a_{12} - a_{11}) 
C_g(2 l' l; 0 0 0) \{ C_g(2 l' l; 1 0 1) \delta_{m 1} - 
C_g(2 l' l; \underline{1} 0\underline{1}) \delta_{m \underline{1}} \} 
\nonumber \\
& + & \frac{1}{3} (a_{23} - a_{21} - a_{22}) \{ \delta_{ll'} \delta_{m0} - 
\left(\frac{2l'+1}{2l+1}\right)^{1/2} 
C_g(2 l' l; 0 0 0)^2 \delta_{m0} - \sqrt{\frac{3}{2}}
\left(\frac{2l'+1}{2l+1}\right)^{1/2} C_g(2 l' l; 0 0 0) \nonumber \\
& & (C_g(2 l' l; 2 0 2) \delta_{m2} +
C_g(2 l' l; \underline{2} 0 \underline{2}) \delta_{m \underline{2}})\}]
\end{eqnarray}

\begin{eqnarray}
\beta C_{1} & = & \frac{\rho_0^2 \sqrt{4 \pi}}{2} \sum_{l_1 l_2 l l'}
\sum_{m_1^{\prime} m_2^{\prime}} \sum_{m \lambda} \sum_{n_1 n_2} \sum_G 
(i)^{l'} \frac{\sqrt{(2l' + 1)}}{(2l_1+1)(2l_2+1)} Q_{l_1 m_1^{'}n_1}(G) 
Q_{l_2 m_2^{'} n_2}(-G) \nonumber \\
& & D_{m\lambda}^{l*}(\hat{\bf n}) J_{l_1 l_2 l l'}^{n_1 n_2} 
[\frac{1}{\sqrt{6}} \left(\frac{2l'+1}{2l+1}\right)^{1/2} (a_{11} - a_{12}) 
C_g(2 l' l; 0 0 0) \{ C_g(2 l' l; 2 0 2) \delta_{m 2} - 
C_g(2 l' l; \underline{2} 0\underline{2}) \delta_{m \underline{2}} \} 
\nonumber \\
& - & \frac{1}{3} \{ \frac{1}{2} (a_{11} + a_{12}) + m_2^{'} 
(a_{11} - a_{12}) \}
\{ \delta_{ll'} \delta_{m0} - \left(\frac{2l'+1}{2l+1}\right)^{1/2} 
C_g(2 l' l; 0 0 0)^2 \delta_{m0} \nonumber \\
& + & \sqrt{\frac{3}{2}} \left (\frac{2l'+1}{2l+1}\right)^{1/2} 
C_g(2 l' l; 0 0 0)(C_g(2 l' l; 2 0 2) \delta_{m2} +
C_g(2 l' l; \underline{2} 0 \underline{2}) \delta_{m \underline{2}})\}]
\end{eqnarray}

\begin{eqnarray}
\beta C_{2} & = & -\frac{\rho_0^2 \sqrt{4 \pi}}{2} \sum_{l_1 l_2 l l'}
\sum_{m_1^{\prime} m_2^{\prime}} \sum_{m \lambda} \sum_{n_1 n_2} \sum_G 
(i)^{l'} \frac{\sqrt{(2l' + 1)}}{(2l_1+1)(2l_2+1)} Q_{l_1 m_1^{'}n_1}(G) 
Q_{l_2 m_2^{'} n_2}(-G) \nonumber \\
& & D_{m\lambda}^{l*}(\hat{\bf n}) J_{l_1 l_2 l l'}^{n_1 n_2} 
[\frac{1}{\sqrt{6}} \left(\frac{2l'+1}{2l+1}\right)^{1/2} (a_{11} - a_{12}) 
C_g(2 l' l; 0 0 0) \{ C_g(2 l' l; 2 0 2) \delta_{m 2} - 
C_g(2 l' l; \underline{2} 0\underline{2}) \delta_{m \underline{2}} \} 
\nonumber \\
& + & \frac{1}{3} \{ \frac{1}{2} (a_{11} + a_{12}) + m_2^{'} 
(a_{11} - a_{12})\} \{ \delta_{ll'} \delta_{m0} - 
\left(\frac{2l'+1}{2l+1}\right)^{1/2} 
C_g(2 l' l; 0 0 0)^2 \delta_{m0} \nonumber \\
& - & \sqrt{\frac{3}{2}} \left(\frac{2l'+1}{2l+1}\right)^{1/2} 
C_g(2 l' l; 0 0 0)(C_g(2 l' l; 2 0 2) \delta_{m2} +
C_g(2 l' l; \underline{2} 0 \underline{2}) \delta_{m \underline{2}})\} 
\nonumber \\
& - & \frac{1}{\sqrt{6}} m_2^{'} C_g(2 l' l; m_1^{'} m_2^{'} \lambda) 
\left(\frac{2l'+1}{2l+1}\right)^{1/2} C_g(2 l' l; 0 0 0) 
\{ C_g(2 l' l; 1 0 1) \delta_{m 1} - 
C_g(2 l' l; \underline{1} 0\underline{1}) \delta_{m \underline{1}} \}] 
\end{eqnarray}

where 
\begin{displaymath}
J_{l_1 l_2 l L}^{n_1 n_2} = \int dr \; r^4 \; j_L(G r) \; 
c(l_1 l_2 l; n_1 n_2; r)
\end{displaymath}
are the structural parameters. A line under a numerical subscript denotes a
negative quantity. Eqs.(3.18)-(3.26) give general expressions for the 
curvature elastic constants of the Sm C and other phases of this class 
in the long wavelength limit. From these results one can obtain expressions 
for the elastic constants of a given phase by using the appropriate 
order parameters corresponding to the phase. 

The structures of smectic phases are characterized by three different class of 
order parameters; (i) orientational, (ii)positional, and (iii) mixed \cite {2}.
In general there can be upto $(2l+1)^2$ orientational and so many mixed order 
parameters of rank $l$ for a given value of the reciprocal lattice vector
${\bf G}$. One uses the symmetries of the phase and of the constituent molecule
to reduce this number. Since $Sm C$ phase is biaxial with point symmetry 
$C_{2h}$, we choose the following orientational order parameters at 
$l=2$ level (molecules are assumed to be cylindrically symmetric).

\begin{mathletters}
\begin{eqnarray}
\bar P_2 & = &  \frac{1}{5}  Q_{20}  = 
\langle P_2(\cos\theta) \rangle \nonumber \\
\omega_1 & = & \frac{1}{5}  Q_{2{\underline 1}} = - \frac{1}{5} Q_{21} =
 \left({\frac{3} {2}}\right)^{\frac {1}{2}} \langle  \sin \theta \cos 
    \theta \cos \phi \rangle  \\
\omega_2 & = & \frac {1}{5}  Q_{2{\underline 2}} =   \frac{1}{5}Q_{22} =
\left({\frac {3}{8}}\right)^{\frac {1}{2}}  \langle \sin^2 
\theta \cos 2 \phi \rangle\\ 
\end{eqnarray}
\end{mathletters}
While $ \bar P_2$ measure the degree of alignment of molecular axis along the 
director ${\bf n}$ which makes a non zero angle with the smectic plane normal, 
$\omega_1$ and $\omega_2$ measure the degree of alignment along the director 
${\bf c}$ and the biaxial ordering. Angular bracket indicates the average 
over the orientational distribution .

 For positional ordering one may choose as in the case of Sm A phase, 
one order parameter corresponding $G = G_z= \frac{2\pi}{d}$ where $d$ is 
the layer spacing in equilibrium. Thus 

\begin{eqnarray}
\mu = Q_{00}(G)=\langle \cos 2\pi{\frac {z}{d}} \rangle
\end{eqnarray}

The coupling between the positional and orientational ordering may 
be described by the order parameters $\tau_1$, $\tau_2$ and $\tau_3$, where 

\begin{mathletters}
\begin{eqnarray}
\tau_1 & = & \frac {1} {5} Q_{20}(G)= \langle \cos \frac {2\pi z} {d} 
P_2(\cos \theta) \rangle \\
\tau_2 & = & \frac {1} {5} Q_{2 {\underline 1}}(G)= -\frac {1} {5} Q_{21} (G) =
(\frac {3} {2})^{\frac {1}{2}} \langle \cos \frac {2\pi z} {d} 
\sin \theta \cos \theta \cos \phi \rangle \\
\tau_3 & = & \frac {1} {5} Q_{2 {\underline 2}}(G)= \frac {1} {5} Q_{22}(G) =
(\frac {3} {8})^{\frac {1}{2}} \langle \cos \frac {2\pi z} {d} 
\sin^2 \theta \cos 2 \phi \rangle
\end{eqnarray}
\end{mathletters}

Using these order parameters we give below explicit expressions for the 
curvature elastic constants for Sm C phase consisting of cylindrically 
symmetric molecules;

\begin{mathletters}
\begin{eqnarray}
B_1 & = & M_1\sin^2 \psi  + M_2\cos \psi  +M_3 \\
B_2 & = & M_4\sin^2\psi- M_2\cos \psi + M_5 \\
B_3 & = & M_6\sin^2\psi+M_7 \\
B_{13} & = & 0 \\
A & = & M_8\sin^2 \psi  +M_9 \\
A_{12} & = & M_{10}\sin^2 \psi  +M_{11} \\
A_{21} & = & M_{12}\sin^2 \psi  +M_{13} \\
C_{1} & = & M_{14}\sin^2 \psi  + M_{15}(1+\cos \psi)^2 +M_{16}(1-\cos \psi)^2 
+M_{17}\sin \psi + M_{18}\sin \psi \cos \psi + M_{19} \\
C_{2} & = & M_{20}\sin^2 \psi +M_{21}(1+\cos \psi)^2 +M_{22}(1-\cos \psi)^2 
-M_{17}\sin \psi + M_{23}\sin \psi \cos \psi + M_{24}
\end{eqnarray}
\end{mathletters}

 Here $\psi$ is the tilt angle. The directors $\hat {\bf n}$ and 
$\hat{\bf c}$ of a uniform phase are assumed to be in the xz plane. 
The expressions for the constants
$M_1-M_{24}$ which involve the order and structural parameters and
are, therefore, temperature and density dependent are given in
Appendix A. In writing these expressions we have taken into account
the fact that for a system of rigid axially symmetric molecules 
$n_1 = n_2 = 0$ and

\begin{equation}
 c(l_1l_2l;00;r) = \frac{1}{4\pi}\sqrt{(2l_1+1)(2l_2+1)} c_{l_1l_2l}(r)
\end{equation}
 
 The structural parameters are connected with harmonic coefficients
$c_{l_1l_2l}$ as 

\begin{eqnarray}
J_{l_1l_2l}(0) & = &  \frac{\rho_0^2}{\sqrt{4\pi}}\int_0^\infty 
 dr r^4 c_{l_1l_2l}(r) \\
{\rm and} \nonumber \\
J_{l_1l_2ll'}& =&  \frac{\rho_0^2}{{\sqrt 4\pi}}\int_0^\infty
 dr r^4 j_{l'}(Gr) c_{l_1l_2l}(r) 
\end{eqnarray} 

 Our results given in Eqs.(3.31) show that for cylindrically symmetric
molecules $B_{13} = 0$. This is, however, not true in general. For 
molecules with  broken cylindrical symmetry $B_{13}$ will be non-zero.
Its magnitude, however, appears to be small compared to $B_1, B_2$ and
$B_3$. Madhusudana and coworkers \cite{8} have found
$\frac{B_{13}}{B_1} \approx 0.3$   

\subsection{Expression for the Elastic constant $\bar{B}$}

As mentioned above the expression for the elastic constant $\bar{B}$  
is derived from the expression of $f_2$ given by Eq.(3.16). Since
Sm C as well as Sm $C^*$ has one dimensional positional ordering in 
addition to orientational ordering we have

\begin{eqnarray}
f_2 & = & -\frac{1}{2} {\rho_0}^2 \sum_G \sum_{l_1 l_2 l} 
\sum_{m'_1 m'_2 m'} \sum_{m\lambda} \sum_{n_1 n_2} 
\frac{Q_{l_1 m'_1 n_1}({\bf G})Q_{l_2 m'_2 n_2}(-{\bf G})}{(2l_1 + 1)
(2l_2 + 1)} C_g(l_1 l_2 l; m'_1 m'\lambda)\nonumber \\
& & D_{m\lambda}^{l*}(\hat{\bf n}) \int d{\bf r}
\; [\exp(i{\bf G}_d z_{12}) - \exp(i{\bf G} z_{12})]\delta_{m'm'_2} 
c(l_1 l_2 l; n_1 n_2; r) Y^*_{lm}(\hat{\bf r})
\end{eqnarray} 

where ${\bf G} = \frac{2\pi k}{d}$ {\rm and} ${\bf G}_d = \frac{2\pi k}{d_e}$;
$d_e$ and $d$ are the interlayer spacing of the distorted and undistorted
Sm C (and  Sm $C^*$) phase. Using eqn.(3.11) we expand above equation
in ascending power of dilation parameter $\epsilon = 
\frac{d_e}{d} - 1 $. The term associated with $\frac{1}{2}\epsilon^2$
defines the bulk elastic constant $\bar{B}$. Comparing Eq.(3.35) and
with Eq.(2.4) we get,

\begin{eqnarray}
\bar{B} & = & \frac{kT}{2} {\rho_0}^2 \sum_G \sum_{l_1 l_2 l}
\sum_{m'_1 m'_2 m'} \sum_{m\lambda} \sum_{n_1 n_2}
\frac{Q_{l_1 m'_1 n_1}({\bf G})Q_{l_2 m'_2 n_2}(-{\bf G})}{(2l_1 + 1)
(2l_2 + 1)} C_g(l_1 l_2 l; m'_1 m'\lambda) \nonumber \\
& & D_{m\lambda}^{l^*}(\hat{\bf n}) {\bf G}^2 \int d{\bf r} \; z_{12}^2
\;\exp(i{\bf G} z) c(l_1 l_2 l; n_1 n_2; r) Y^*_{lm}(\hat{\bf r})
\end{eqnarray} 

Using Eq.(3.18) we have 
\begin{eqnarray}
\bar{B} & = & \frac{kT\rho_0^2}{6} \sqrt{4\pi} \sum_G \sum_{l_1 l_2 l l'}
\sum_{m'_1 m'_2 m'} \sum_{\lambda} \sum_{n_1 n_2} (i)^{l'}
\frac{Q_{l_1 m'_1 n_1}({\bf G})Q_{l_2 m'_2 n_2}(-{\bf G})}
{(2l_1 + 1)(2l_2 + 1)} \nonumber \\
& & {\bf G}^2 \; (2l' + 1)^{1/2}\; D_{0\lambda}^{l^*}(\hat{\bf n}) 
\;C_g(l_1 l_2 l; m'_1 m'\lambda) J_{l_1 l_2 l l'}^{n_1 n_2}
\left[\delta_{ll'} + 2 \left(\frac{2l' +1}{2l +1}\right)^{1/2} 
C_g(2l' l; 000)^2\right]
\end{eqnarray}

As shown in I, the term proportional to $\epsilon$ in the free energy expansion
gives the condition for the interlayer spacing. In terms of the order
parameters given by Eqs.(3.29)-3.30), Eqs.(3.37) can be written as 

\begin{equation}
\beta \bar B = L_1 \sin^2 \psi + L_2 \sin\psi \cos\psi + L_3
\end{equation}

 The expressions for $L_1$, $L_2$ and $L_3$ for a system
consisting of cylindrically symmetric molecules are given in Appendix B.  

\section{Discussions}

 In order to estimate the values of elastic constants given above
we need the values of the order parameters, generalised spherical
harmonic coefficients of the direct pair correlation functions of
an effective fluid as a function of temperature and density and the 
information about the constituent molecules, {\it viz.} electric
multipole moments, geometry of the repulsive core,length-to-width 
ratio, etc. as input parameter.

 Since in the limit of long wavelength distortions the magnitude
of the order parameters are assumed to remain unchanged, the 
value of the order parameters at a given temperature and the
density may be taken either determined experimentally or calculated
from a theory.
 The $c$-harmonics for a given system can in principle, be found either by 
solving the Ornstein-Zernike equation with suitable closure relations
\cite{6} or by adopting a perturbative scheme which is based on the fact that
the fluid structure at high densities is primarily controlled by the 
repulsive part of the interactions. However, such calculations for non-axial
molecules are very complicated and may need enormous computational efforts
to generate reliable data for $c$ harmonics \cite {9}. 

 The difficulty which arises in applying the theory to real systems is 
related to the potential energy of interaction between mesogenic molecules. 
The mesogenic molecules are large and have groups of atoms with their 
own local features.
One way to construct the potential energy of interaction between two such
molecules is to sum the interatomic or site-site potential between atoms
or between interaction sites. However, for mesogenic molecules there are too
many terms in this sum to be practical. Moreover, the dependence of 
interaction on molecular orientations in this expression is implicit 
so that it is difficult to use it in the calculation of angular 
orientation which give rise to liquid crystals.  

In another and more convenient approach one uses rigid molecular 
approximation in which it is assumed that the intermolecular potential 
energy depends upon only on the position of the centre of mass and on 
their orientations. This kind of approach neglects the flexibility of 
molecular structure which plays important role in stability of many 
liquid crystalline phases. In view of various complexities in the 
intermolecular interaction one is often forced to use a phenomenological 
description, either as a straight forward model 
unrelated to any particular physical system, or as a basis for describing
by adjustable parameters fitted to experimental data for interaction 
between  two molecules. Most commonly used models are hard-ellipsoids of 
revolution, hard sphero-cylinders, cut-sphere \cite {10} and Gay-Berne 
\cite {11}. However, non of these are known to show the existence of Sm C 
phase.

 The computer simulation study \cite{12} shows that the GB model exhibits 
Sm A and Sm B phases. The GB potential contains four independent 
parameters which control the anisotropy in the attractive
and repulsive interactions and can be written as  

\begin{equation}
u(1,2) = 4 \epsilon(\hat{\bf r},{\bf \Omega_1},{\bf \Omega_2}) 
\left [ {\{ \frac{\sigma_0} {r - \sigma(\hat{\bf r},{\bf \Omega_1},
{\bf \Omega_2}) + \sigma_0} \} }^{12} - {\{ \frac{\sigma_0}
{r - \sigma(\hat{\bf r},{\bf \Omega_1},{\bf \Omega_2}) + 
\sigma_0} \} }^6 \right ]
\end{equation}
Here $\epsilon(\hat{\bf r},{\bf \Omega_1},{\bf \Omega_2})$ and 
$\sigma(\hat{\bf r},{\bf \Omega_1},{\bf \Omega_2})$ are angle 
dependent strength and range parameters, respectively, and
are defined as 

\begin{eqnarray}
\epsilon(\hat{\bf r},{\bf \Omega_1},{\bf \Omega_2}) & = & 
\epsilon_0 [1 - \chi^2 (\hat{\bf e_1}. \hat{\bf e_2})^2]^{-1/2} 
\left [ 1 - \chi^{\prime} \frac{(\hat{\bf r}.\hat{\bf e_1})^2
+ (\hat{\bf r}.\hat{\bf e_2})^2 - 2 \chi^{\prime} (\hat{\bf r}.
\hat{\bf e_1})(\hat{\bf r}.\hat{\bf e_2})(\hat{\bf e_1}.\hat{\bf e_2})}
{1 - {\chi^{\prime}}^2(\hat{\bf e_1}.\hat {\bf e_2})^2}  \right ]^2 \nonumber \\
\sigma(\hat{\bf r},{\bf \Omega_1},{\bf \Omega_2}) & = & 
\sigma_0 \left [ 1 - \chi
\frac{(\hat{\bf r}.\hat{\bf e_1})^2 + (\hat{\bf r}.\hat{\bf e_2})^2 - 2 \chi 
(\hat{\bf r}.\hat{\bf e_1})(\hat{\bf r}.\hat{\bf e_2})
(\hat {\bf e_1}.\hat{\bf e_2})}{1 - \chi^2 (\hat{\bf e_1}.\hat{\bf e_2})^2}
 \right ]^{-1/2}  
\end{eqnarray}
where $\hat{\bf e_1}$ and $\hat{\bf e_2}$ are the unit vectors along the 
symmetry axes of two interacting molecules. $\epsilon_0$ and 
$\sigma_0$ are parameters which provide a measure of the attractive 
interactions and the width of the molecule. The anisotropy 
parameter $\chi$ and $\chi^{\prime}$ are defined as 
\begin{displaymath}
\chi = \frac{x_0^2 - 1}{x_0^2 + 1} \; \; {\rm and} \; \; 
\chi^{\prime}  = \frac{k'^{1/2} - 1}{k'^{1/2} + 1} 
\end{displaymath}
where $x_0 = 2a/2b$ is the length (major axis) to breadth (minor
axis) ratio and $k'$ is the ratio of the potential well depth for
the side by side and end to end configurations. We have taken 
here $x_0 = 3.0$ and $k' = 5$. The harmonics were generated 
by solving the Ornstein-Zernike equation using the 
Percus-Yevick closure relation
\cite{13}. Using these $c$-harmonics we have calculated the values
of the structural parameters at $k_BT/\epsilon_0 = 0.8$ and the packing
fraction $\eta = (\frac{\pi}{6}) x_0 \rho_0 \sigma_0^3 = 0.49$.
The values of few structural parameters for $G=0$ and 
$G=\frac{2\pi}{x_0}= 2.0944 $ so obtained are as follows

\begin{eqnarray}
J_{220}(0)  =  2.23174 &\;\;\;& J_{0000} = -2.78902 \nonumber \\
J_{222}(0)  =  0.33324 &\;\;\;& J_{0002}  = -4.61192 \nonumber \\
J_{224}(0)  = -4.02296 &\;\;\;& J_{2020}  =  0.07539 \nonumber \\
J_{242}(0)  =  0.30951 &\;\;\;& J_{2022}  = -2.05825 \nonumber \\
J_{244}(0)  = -1.06209 &\;\;\;& J_{2024}  = -0.56503 \nonumber \\
J_{440}(0)  =  0.49893 &\;\;\;& J_{2200}  =  0.13322 \nonumber \\
J_{442}(0)  = -0.09114 &\;\;\;& J_{2202}  =  0.6818 \nonumber \\
J_{444}(0)  =  0.13077 &\;\;\;& J_{2220} =  -0.49332 \nonumber \\ 
 &\;\;\;& J_{2222}  =   - 0.00627  \nonumber \\
 &\;\;\;& J_{2224}  =    0.17776  \nonumber \\
 &\;\;\;& J_{2240}  =    0.54022  \nonumber \\
 &\;\;\;& J_{2420} =  - 0.25849  \nonumber 
\end{eqnarray}

 Since the values of the order parameters are not known, we assume
them to be equal with value 0.5 to estimate the relative 
contributions made by different terms in Eqs. (3.31) and (3.38). 
It is found that
(i) dominant contributions for all elastic constants come from 
angle independent term, (ii) coefficients of linear terms {\it i.e.}
terms involving $\sin \psi$ or $\cos \psi$ are small compared 
to terms involving $\sin^2\psi$and
(iii) the relative contributions to $A$, $A_{12}$, $A_{21}$ 
made by angle dependent terms  are small compared to their relative
contributions to $B_1$, $B_2$ and  $B_3$. The values of all these constants
except  $B_{13}$ are found to be of the order of $10^{-6} $ dynes.
As has already been pointed out that for linear rigid molecule $B_{13}$
is zero. For molecules with broken axial symmetry  $B_{13}$ is non-zero
but its value is less compared to other elastic constants.  
 The elastic constant ${\bar B}$ is found to be of same order as in 
Sm A. However, as tilt angle increases the value of  ${\bar B}$ 
decreases. The two angle dependent terms in Eqs.(3.38) is found to 
cancel each other, and therefore, the major contribution is due to
the angle independent term.

\acknowledgements

The work was supported by the Department of Science 
and Technology (India) through project grant.

\newpage

\begin{center}
 {\bf APPENDIX A}
\end{center}

 In this appendix we give expilicit expressions for the 
coefficients $M_1$ to $M_{24}$ in units of $k_BT/\sigma_0$ 
(see Eqs.(3.31)).
As is obivious from Eqs. (3.19)-(3.27), they involve order
and structural parameters of all orders. We list only first few
non-vanishing terms.   

\begin{eqnarray}
M_1 & = & -\omega_1^2 \left[ \frac{1}{2} \sqrt{\frac{5}{14}}
J_{222}(0)\right] + \omega_2^2 \left[ 6\sqrt{\frac{10}{7}} J_{222}(0)
\right] +  {\bar P_2}\omega_2 \left[-\frac{4}{3} 
\sqrt{\frac{15}{7}} J_{222}(0) \right]+..... \nonumber\\
M_2 & = & \omega_1^2 \left[ 2 {\sqrt\frac{5}{14}} J_{222}(0)\right] 
- {\bar P_2}\omega_2 \left[4 {\sqrt\frac{5}{21}} J_{222}(0) \right]+
..... \nonumber\\
M_3 & = & \omega_1^2 \left[ \frac{1}{3} {\sqrt\frac{5}{14}}
(-2 J_{222}(0)+{\sqrt{14}}J_{220}(0))\right] +  
\omega_2^2 \left[ 4\frac{{\sqrt 5}}{3}(J_{220}(0)- 
{\sqrt\frac{2}{7}}J_{222}(0))
\right]+ \nonumber\\
& & + {\bar P_2}\omega_2 \left[4{\sqrt\frac{5}{21}} J_{222}(0) \right]+ 
.....\nonumber \\
M_4 & = & -\tau_1\tau_3\left[ \frac{30}{7}{\sqrt \frac{10}{7}}
 J_{2222}\right]+..... \nonumber\\ 
M_5 & = & \omega_1^2 \left[\frac{1}{3} {\sqrt \frac{5}{14}}
(4J_{222}(0)+{\sqrt{14}} J_{220}(0)\right] +
\omega_2^2 \left[ 4\frac{{\sqrt 5}}{3}(J_{220}(0)- {\sqrt \frac{2}{7}}
(J_{222}(0))\right] \nonumber \\
& & -{\bar P_2}\omega_2 \left[4{\sqrt \frac{5}{21}}
 J_{222}(0) \right]+.....  \nonumber\\ 
M_6 & = & -\omega_2^2 \left[ 4 {\sqrt\frac{10}{7}} J_{222}(0)\right]
+ {\bar P_2}\omega_2 \left[\frac{4}{3} 
{\sqrt\frac{15}{7}} J_{222}(0) \right] + .....\nonumber\\ 
M_7 & = & -\omega_1^2 \left[ \frac{1}{3}{\sqrt\frac{10}{7}}
(J_{222}(0)-{\sqrt\frac{7}{2}}J_{220}(0)) \right]
 + \omega_2^2 \left[ 4\frac{\sqrt 5}{3}( J_{220}(0)
+ 2{\sqrt\frac{2}{7}} J_{222}(0)\right] +..... \nonumber  \\
M_8 & = & {-\bar P^2_2} \left[2{\sqrt \frac{10}{7}}J_{222}(0) \right]
+.....\nonumber\\ 
M_9 & = & \omega_1^2 \left[ -\frac{2}{3} \sqrt{\frac{5}{14}}
J_{222}(0)+ \frac{\sqrt{5}}{3}J_{220}(0) \right] 
{-\bar P^2_2} \left[\frac{3\sqrt{5}}{2}( J_{220}(0) -\frac{2}{3}
\sqrt{\frac{2}{7}}J_{222}(0))\right] +.....\nonumber\\
M_{10} & = &  {-\bar P^2_2} \left[{\sqrt \frac{10}{7}}
J_{222}(0) \right] +.....\nonumber\\ 
M_{11} & = & \omega_1^2 \left[ \frac{4}{21} {\sqrt\frac{5}{14}}
(-14 J_{222}(0)+\frac{35{\sqrt{14}}}{8}J_{220}(0))\right] + 
{\bar P^2_2} \left[\frac{{\sqrt 5}}{2} J_{220}(0) \right] +.....\nonumber\\
M_{12} & = & -\tau_1^2 \left[\frac{15}{7}
{\sqrt \frac{10}{7}}J_{2222} \right]+ .....\nonumber\\
M_{13} & = & \omega_1^2 \left[ \frac{1}{21} {\sqrt \frac{5}{14}}
(28 J_{222}(0) +  7{\sqrt{14}} J_{220}(0))\right]
 + {\bar P^2_2} \left[\frac{{\sqrt 5}}{2} J_{220}(0)\right] +.....\nonumber\\ 
M_{14} & = & -{\bar P_2}\omega_1 \left[\frac{35}{8} 
{\sqrt\frac{3}{35}} J_{222}(0) \right] +.....\nonumber\\
M_{15} & = & {\bar P_2}\omega_1 \left[\frac{25}{16{\sqrt {105}}} 
 J_{222}(0) \right] +.....\nonumber\\
M_{16} & = & {\bar P_2}\omega_1 \left[\frac{5}{16{\sqrt {105}}} 
J_{222}(0) \right] +.....\nonumber\\
M_{17} & = & {-\bar P^2_2} \left[\frac{1}{2}{\sqrt \frac{5}{14}} 
J_{222}(0) \right] +.....\nonumber\\
M_{18} & = & {\bar P^2_2} \left[\frac{1}{2}{\sqrt \frac{5}{14}} 
J_{222}(0) \right] +.....\nonumber\\
M_{19} & = &  {\bar P_2}\omega_1 \left[\frac{5}{28}{\sqrt \frac{3}{35}}
(7 J_{222}(0)+\frac{28}{3}{\sqrt \frac{7}{2}}J_{220}(0)) \right] + 
.....\nonumber\\ 
M_{20} & = & {-\bar P_2} \omega_1\left[\frac{5}{8}{\sqrt \frac{3}{35}} 
J_{222}(0) \right] +.....\nonumber\\
M_{21} & = & {-\bar P_2} \omega_1\left[\frac{25}{48}{\sqrt \frac{3}{35}} 
J_{222}(0) \right] +.....\nonumber\\
M_{22} & = & {-\bar P_2} \omega_1\left[\frac{5}{48}{\sqrt \frac{3}{35}} 
J_{222}(0) \right] +.....\nonumber\\
M_{23} & = & \tau_1^2 \left[\frac{15}{28}
{\sqrt \frac{5}{14}}J_{2222} \right]+.....\nonumber\\
M_{24} & = & {\bar P_2} \omega_1\left[\frac{5}{28}{\sqrt \frac{3}{35}} 
(7 J_{222}(0) + \frac{14{\sqrt{14}}}{3} J_{220}(0))\right] +.....\nonumber
\end{eqnarray}

\newpage
\begin{center}
{\bf APPENDIX B}
\end{center}
 
 We give explicit expressions for the coefficients $L_1-L_3$
involving first few terms of order and structural parameters.
 
\begin{eqnarray}
L_1 & = & \frac{1}{6} \sum_G \left[(\frac{{\sqrt 6}}{2}-3)G^2\mu\tau_1
I_{022} + {\sqrt \frac{5}{7}}(\frac{3}{\sqrt 2} + {\sqrt 3})
G^2\tau^2_1 I_{222} + .....\right] \nonumber \\
L_2 & = & \frac{1}{6} \sum_G \left[{\sqrt 6}G^2\mu\tau_2
I_{022} + {\sqrt \frac{5}{7}}(3{\sqrt 2} - {\sqrt 3})
G^2 \tau_1 \tau_2 I_{222} + ...... \right] \nonumber \\ 
L_3 & = & \frac{1}{6} \sum_G \left[\mu^2 G^2 I_{000} + 
2G^2 \mu\tau_1 I_{022} + {\sqrt 5}G^2 \tau_1^2 I_{220}
- {\sqrt\frac{10}{7}} G^2\tau_1^2 I_{222} + ..... \right] \nonumber \\ 
{\rm where} \nonumber \\
I_{000} & = & J_{0000} - 2J_{0002} \nonumber \\ 
I_{022} &= & 2J_{0220} - \frac{55}{7}J_{0222} + \frac{36}{7}J_{0224}\nonumber \\
I_{220} &= & J_{2200} - 2J_{2202}\nonumber \\
I_{222} &= & 2J_{2220} - \frac{55}{7}J_{2222} + \frac{36}{7}J_{2224} \nonumber
\end{eqnarray}

\newpage

\begin{figure}
\caption{Molecular arrangements in Sm C, the directors $\hat{\bf n}$, 
$\hat{\bf c}$ and the space-fixed coordinate frame}
\end{figure}
\end{document}